\documentclass[aps,prb,twocolumn,floats,tightenlines,floatfix,groupedaddress]{revtex4}
\usepackage{amsmath}

\begin{document}

%\preprint{\tt Report No.XXXXXX}

\title{Application of Pad\'{e} interpolation to stationary state
problems}
\author{C. N. Leung}\email{leung@physics.udel.edu}
\author{Yvonne Y. Y.
Wong}\email{ywong@physics.udel.edu}\affiliation{ Department of
Physics and Astronomy, University of Delaware, Newark, Delaware
19716}

%\date{\today}
%\vspace*{1.0 cm}

\begin{abstract}
If the small and large coupling behavior of a physical system can
be computed perturbatively and expressed respectively as power
series in a coupling parameter $g$ and $1/g$, a Pad\'{e}
approximant embracing the two series can interpolate between these
two limits and provide an accurate estimate of the system's
behavior in the generally intractable intermediate coupling
regime. The methodology and validity of this approach is
illustrated by considering several stationary state problems in
quantum mechanics.
\end{abstract}

\maketitle

\section{Introduction}
A Pad\'{e} approximation is a formal transformation of the first
$n$ terms in the power series for a function $f(x)$ into a
rational function. The rational function $R(x)$, known as the
Pad\'{e} approximant, typically comprises a ratio of two
polynomials of $x$, chosen in a way that a Taylor expansion of
$R(x)$ completely reproduces the original power series up to order
$n$. In contrast to the truncated ``mother'' series,
the Pad\'{e} approximant is able to maintain
remarkable fidelity to the true $f(x)$ for values of $x$ well
beyond the radius of convergence of the original power series,
although how accurate it is, or how large $|x|$ may be before the
approximation fails is impossible to establish in general.
Nevertheless, its extraordinary predictive power has in the past
been exploited in areas of quantum field theory and statistical
physics. In quantum electrodynamics and quantum chromodynamics,
for example, the Pad\'{e} method has been shown to be an effective
means of both estimating unknown higher order terms as well as of
summing the perturbation series for a physical observable that has
been calculated to some finite order in the coupling
constant.\cite{bib:ellis}

In a recent paper,\cite{bib:quarkonium} one of us has explored a
different usage
of the Pad\'{e} approximation in which the Pad\'{e}
approximant, constructed from the truncated power series of $f(x)$
about two different points, serves to interpolate $f(x)$ between
the two points of expansion. In particular, if one is able to
compute perturbatively both the small $x$ and large $x$ behavior
of $f(x)$ ($x > 0$), and express them respectively in a power
series in $x$ and $1/x$, a Pad\'{e} approximant that
simultaneously satisfies the two perturbation series will provide
an accurate estimation of $f(x)$ for the entire range of $x$,
provided that $f(x)$ is a sufficiently smooth function in this
range. This approach is known as the two-point Pad\'{e}
approximation,\cite{bib:2ptPade} and is especially savoring from
the perspective of duality in supersymmetric gauge
theories;\cite{bib:duality}
if one can infer perturbatively the strong coupling behavior of a
theory from its weakly coupled dual theory, a Pad\'{e}
interpolation of this limit with the perturbative weak coupling
limit of the original theory will then give us a handle on the
behavior of the theory for all coupling strengths.

The Pad\'{e} interpolation method may also be employed in some
cases where there are no obvious expansion parameters. An example
is a system for which the Hamiltonian $H$, although itself not
exactly soluble, can be expressed as a sum of two constituent
Hamiltonians, $H_1$ and $H_2$, both with known exact solutions. It
was proposed in Ref.~\onlinecite{bib:quarkonium} that, by tagging
either $H_1$ or $H_2$ with an interpolation parameter $\lambda$, the
two required perturbative series in opposite limits of $\lambda$ could
be generated, and an estimate for $H$ would correspond to evaluating
the Pad\'{e} approximant with $\lambda=1$. This method was applied
in Ref.~\onlinecite{bib:quarkonium} to heavy quarkonium systems
with reasonable success.

The quarkonium example demonstrates the power and usefulness of
Pad\'{e} interpolation for treating a certain class of stationary
state problems. The technique may serve as an improvement or a
supplement to perturbation theory typically taught in courses on
quantum mechanics. The purpose of this note is to present a
pedagogical discourse of the methodology of Pad\'{e} interpolation.
For the examples considered below, the Pad\'{e} interpolation method
is shown to be stable to higher orders, and yield particularly good
results when the usual perturbative method fails.

\section{The Method}\label{sec2}
Consider a system governed by a Hamiltonian $H$ that has no
known solution, but which may be expressed as a sum of two
parts: $H = H_1 + H_2$, where $H_1$ and $H_2$ are
{\it individually} exactly soluble. We are interested in
finding the spectrum of $H$. In order to implement the
Pad\'{e} interpolation for this purpose, we introduce the
interpolation Hamiltonian
\begin{equation}
\label{eq:interpolatingH}
H(\lambda) \equiv H_1 + \lambda H_2\,,
\end{equation}
where the interpolation parameter $\lambda$ is real and positive.
Suppose first that $\lambda \ll 1$. We may then regard $H_2$
as a perturbation, and calculate the eigenvalues of $H$ as power
series in $\lambda$ (the subscript $j$ labels the eigenvalues):
\begin{equation}
\label{eq:smallx}
E_j^<(\lambda) = c_0 +c_1 \lambda + c_2 \lambda^2
+ \ldots + c_m \lambda^m\,.
\end{equation}
In the opposite $\lambda \gg 1$ limit, we rewrite
Eq.~(\ref{eq:interpolatingH}) as
\begin{equation}
H'(\lambda) = \lambda \left[H_2 + \left(\frac{1}{\lambda} \right)
H_1 \right]\,,
\end{equation}
and a second perturbative solution of $H(\lambda)$ follows from
treating $1/\lambda$ as a small parameter:
\begin{equation}
\label{eq:largex} E_j^>(\lambda) = \lambda \left[ b_0 + b_1
\left(\frac{1}{\lambda}\right) + b_2 \left(\frac{1}{\lambda}
\right)^2 + \ldots + b_n \left(\frac{1}{\lambda} \right)^n
\right].
\end{equation}
Note that both $H_1$ and $H_2$ must admit stationary states and
be able to be treated as perturbations for the method to work.
An example of a Hamiltonian that admits bound-state solutions
but that cannot be treated as a perturbation is the Hamiltonian for
an infinite rectangular potential well.

A generic Pad\'{e} approximant for the energy eigenvalues has the
form
\begin{equation}
E_j^{\rm PA}(\lambda) = \frac{p_0 + p_1 \lambda + p_2 \lambda^2 +
\ldots + p_N \lambda^N}{1 + q_1 \lambda + q_2 \lambda^2 + \ldots +
q_{M} \lambda^{M}}\,,
\end{equation}
where the $N+M+1$ coefficients are determined by matching order by
order the power series expansion of the Pad\'{e} approximant with
the perturbative results. For $b_0 \neq 0$, the nature of
$E_j^>(\lambda)$ in Eq.~(\ref{eq:largex}) demands that the
polynomials in the numerator and the denominator of the Pad\'{e}
approximant differ by one degree such that $M=N-1$. Furthermore,
suppose that we solve $H(\lambda)$ for small and large $\lambda$
to the same order in perturbation theory, that is, $m=n$. (This
case is just an illustration and is not a necessary condition for
implementing the Pad\'{e} interpolation, although the accuracy of
the approximation will depend on $m$ and $n$.) Then
Eqs.~(\ref{eq:smallx}) and (\ref{eq:largex}) together furnish
$2n+2$ simultaneous equations for the $2N$ unknown coefficients
$p$ and $q$, and consequently the polynomials in the Pad\'{e}
approximant must satisfy $N=n+1$ and $M=n$.
The final step of setting $\lambda=1$ in the Pad\'{e} approximant
yields an estimate for the eigenvalues of the original Hamiltonian
$H$.

\section{Examples}\label{sec3}
We demonstrate here the validity of the Pad\'{e} interpolation
method by way of two examples. Consider first a simple two-state
system described by the Hamiltonian
\begin{equation}
\label{eq:2stateH}
H = a \sigma_y + b \sigma_z\,,
\end{equation}
where $a$ and $b$ are real parameters, and $\sigma_y$ and
$\sigma_z$ are the Pauli matrices. For example, this
Hamiltonian can represent the interaction energy of a charged
spin-1/2 particle in a magnetic field $\vec{B} = (0, B_y, B_z)$.
In this case,
\begin{equation}
\label{eq:Bparameters}
a = - \frac{g \hbar B_y}{2} ~~~{\rm and}~~~ b = - \frac{g \hbar
B_z}{2}\,,
\end{equation}
where $g$ is the gyromagnetic ratio of the particle. This
example is trivial in the sense that $H$ can be easily
diagonalized to yield the exact eigenvalues
\begin{equation}
\label{eq:exactE}
E_{\pm} = \pm \sqrt{a^2 + b^2}\,.
\end{equation}
However, the comparison of these exact results with the approximate
eigenvalues obtained below by Pad\'{e} interpolation will
provide a way to gauge the accuracy of the approximation method.

For $|a| \gg |b|$, for example, if the magnetic field is almost
aligned with the $y$-axis, the $b \sigma_z$ term in $H$ may be
treated as a perturbation. To be more explicit, we may express the
Hamiltonian as
\begin{equation}
\label{eq:Ha>>b}
H^{|a| \gg |b|} = a \left(\sigma_y + \frac{b}{a} \sigma_z\right),
\end{equation}
which has the form of Eq.~(\ref{eq:interpolatingH}), except that
$\lambda$ is equal to $b/a$ and corresponds to a physical
expansion parameter. If we calculate the energy eigenvalues to
second order in perturbation theory, we find
\begin{equation}
 E^{|a| \gg |b|}_{\pm} = \pm |a| \left(1 +
\frac{b^2}{2a^2}\right).
\label{E}
\end{equation}
For $|a| \ll |b|$, for example,
if the magnetic field is
almost parallel to the $z$-axis, the $a \sigma_y$ term in $H$ can be
regarded as a perturbation. We find that, again to second order
in perturbation theory (with $a/b$ as the expansion parameter), the
eigenvalues of $H$ are now given by
\begin{equation} E^{|b| \gg |a|}_{\pm} = \pm |b| \left(1 +
\frac{a^2}{2b^2}\right).
\label{E'}
\end{equation}

A Pad\'{e} approximant that interpolates these two limits of the
energy eigenvalues can now be constructed. For the eigenvalue
$E_+$, we find
\begin{equation} E_{+}^{\rm PA} = |a| \frac{|b/a|^3 + \frac{3}{2}
|b/a|^2 +
\frac{3}{2} |b/a| + 1}{|b/a|^2 + \frac{3}{2} |b/a| + 1}\,.
\label{EPA}
\end{equation}
This Pad\'{e} approximant is uniquely determined from the
perturbative expansions for $E_+$ given in Eqs.~(\ref{E}) and
(\ref{E'}). Table~\ref{table:2state} compares this Pad\'{e}
interpolation result with the exact eigenvalue,
Eq.~(\ref{eq:exactE}), for various values of the parameter $|b/a|$.
We see that the Pad\'{e} interpolation yields an approximation
that is within 1\% of the exact result for all values of
$|b/a|$. This simple example demonstrates the potential power of
the Pad\'{e} interpolation technique: by simply computing the
leading perturbative corrections for small and large $|b/a|$, one
obtains a very accurate approximation to the eigenvalues for all
values of $|b/a|$.

As a second example, consider a single particle subject to a
one-dimensional linear plus harmonic oscillator potential.
(The Pad\'{e} interpolation technique we shall use to solve
this problem is similar to that applied in
Ref.~\onlinecite{bib:quarkonium} to nonrelativistic quarkonium
systems.) The Hamiltonian that describes the
motion of the particle is
\begin{equation}
\label{eq:exampleham}
H = - \frac{\hbar^2}{2m} \frac{d^2}{dx^2} + \gamma x +
\frac{1}{2}m \omega^2 x^2 + V(x)\,,
\end{equation}
where $\gamma > 0$. (The latter condition is necessary in order for
the Hamiltonian $H_1$ in Eq.~(\ref{eq:h1}) to admit stationary state
solutions.) $V(x)$ represents the rigid wall potential:
\begin{equation}
V(x)=
\begin{cases}
0 & \text{$x>0$} \\
\infty\,. & \text{otherwise}
\end{cases}
\end{equation}
The presence of $V(x)$ restricts the particle's motion to be along
the positive $x$-axis.

The Hamiltonian in Eq.~(\ref{eq:exampleham}) without $V(x)$ appears
in many textbooks on quantum mechanics,\cite{bib:Sakurai} and can
be easily solved by transforming to a new coordinate $x'$ with the
origin at $x = -
\gamma/(m \omega^2)$. The presence of the rigid wall potential,
however, requires all wave functions to vanish for $x \le 0$ and
renders such a coordinate redefinition useless. The Hamiltonian with
$\gamma = 0$ is also a typical textbook problem\cite{bib:Griffiths}
that is exactly soluble; the boundary condition due to the rigid
wall forces all energy eigenfunctions to vanish at the origin,
which implies that only the harmonic oscillator states with odd
parity are allowed.

Solving the complete Hamiltonian (\ref{eq:exampleham}) is a
somewhat more challenging task. In particular, if the linear and
quadratic potentials are comparable, conventional perturbative
methods are not applicable. We shall therefore resort to the method
outlined in Sec.~\ref{sec2} to find its eigenvalues. Note
that as long as we confine the particle's motion to the positive
$x$ branch, and impose the boundary condition that all
eigenfunctions vanish at $x = 0$, we may drop the rigid wall
potential $V(x)$ in Eq.~(\ref{eq:exampleham}). The resulting
Hamiltonian can then be cast in the form of
Eq.~(\ref{eq:interpolatingH}), with
\begin{eqnarray}
\label{eq:h1}
H_1 &=& -\frac{1}{4}\frac{d^2}{dx^2} + \gamma x\,,\\
\noalign{\noindent and}
\label{eq:h2}
H_2 &=& -\frac{1}{4}\frac{d^2}{dx^2} + \frac{1}{2} x^2\,,
\end{eqnarray}
where, for simplicity, we have set $\hbar=1$, $m=1$ and $\omega=1$
such that the arbitrary parameter $\gamma$ alone regulates the
relative importance of the two potential energy terms. Note that
it is also necessary to split the kinetic energy term. Here, we
have arbitrarily put half of the original kinetic energy term into
each of the sub-Hamiltonians in Eqs.~(\ref{eq:h1}) and
(\ref{eq:h2}). As we will see (in the last
paragraph of this
section), better accuracy will generally be achieved in Pad\'{e}
interpolation if a larger fraction of the kinetic energy is
included in the sub-Hamiltonian containing the dominant potential
energy term.

The solutions to $H_1$ are the familiar Airy functions $Ai(z)$, with
$z=(2/\gamma)^{2/3} (\gamma x - \epsilon_j^<)$; the energy
eigenvalues $\epsilon_j^<$ are determined by the roots of $Ai(z)$.
On the other hand, $H_2$ is solved by $\exp(-\xi^2/2) h_j(\xi)$,
where $h_j(\xi)$ are the Hermite polynomials of degree $j$,
$\xi = 2^{1/4} x$, $\epsilon_j^> =(j + 1/2)/\sqrt{2}$ are the
allowed energies, and the index $j$ must be an odd integer in
order to satisfy the boundary condition $\psi_j(0) = 0$, where
$\psi_j(x)$ denotes the stationary state wave functions.

We now proceed to perform the relevant perturbative
calculations. We have evaluated to first, second, and third order
in $\lambda$ and $1/\lambda$ the approximate ground and first
excited state energies for $\gamma=1$, that is, when the linear and
quadratic potential energy terms are comparable, and have formed the
unique Pad\'{e} approximant for each instance. Because closed-form
expressions for integrals involving Airy functions generally do
not exist, we did the exercise numerically. As an
illustration, the analysis of the ground state generates the two
series,
%\begin{subequations}
\begin{eqnarray}
E_0^< & =& 1.47292 + 1.06950 \lambda - 0.0131354 \lambda^2 +
\ldots, \label{eq:smseries}\\ E_0^> &=& 1.06006 \lambda + 1.47918
- 0.00467253 \frac{1}{\lambda} + \ldots. \label{eq:lgseries}
\end{eqnarray}
%\end{subequations}
The first (second) order Pad\'{e} approximant,
\begin{eqnarray}
E^{\rm PA}_{0,{\rm 1st}} &=& \! \frac{1.47292 + 3.14779 \lambda +
1.49659 \lambda^2}{1+1.41100 \lambda}, \\ E^{\rm PA}_{0, {\rm
2nd}} &=& \! \frac{1.47292 + 5.36462 \lambda + 6.07765 \lambda^2 +
2.14021 \lambda^3}{1 + 2.91607 \lambda + 2.01781
\lambda^2},\nonumber \\
\end{eqnarray}
follows from manipulating the first two (three) terms of
Eqs.~(\ref{eq:smseries}) and (\ref{eq:lgseries}).
Tables~\ref{table:gamma1} and \ref{table:gamma2} contain a summary
of the results for the ground and first excited states.

It is instructive to compare these results with those one
would obtain from conventional perturbative calculations alone.
Because there is no preference for either of the two potential
energy terms, we consider both of the following parameterizations
of the Hamiltonian:
\begin{eqnarray}
\label{eq:alpha} H(\alpha)&=&-\frac{1}{2}\frac{d^2}{dx^2} +
\frac{1}{2}x^2 + \alpha \gamma x + V(x)\,, \\
\noalign{\noindent and}
\label{eq:beta} H(\beta)&=&-\frac{1}{2}\frac{d^2}{dx^2} + \gamma x
+
\beta \frac{1}{2}x^2 + V(x)\,,
\end{eqnarray}
where $\alpha$ and $\beta$ are the small parameters that are
eventually set to 1. As seen in Tables~\ref{table:gamma1} and
\ref{table:gamma2}, the Pad\'{e} interpolation gives by far the
most stable results.
A further comparison with exact
solutions from the numerical integration of the Schr\"{o}dinger
equation, also given in Tables~\ref{table:gamma1} and
\ref{table:gamma2}, exemplifies the validity of the method. We
have also checked the accuracy of the method for higher excited
states. The approximate energies obtained, even to first order in
the perturbation parameters, are always accurate to within 1\% of
their exact values.

For completeness, we have examined situations in which one
potential energy term is dominant, and perturbative calculations
on the smaller term alone are expected to yield reasonably
accurate results. This is certainly the case. However, as seen in
Tables~\ref{table:ho} and \ref{table:af}, the Pad\'{e}
interpolation is able to do a better job, provided that the
original kinetic energy term is distributed among the two
sub-Hamiltonians (\ref{eq:h1}) and (\ref{eq:h2}) in a way that
reflects the relative significance of the two potential energy
terms. We have also studied the effects of distributing the
kinetic energy unevenly between the two sub-Hamiltonians in the
$\gamma = 1$ case. As shown in Tables~\ref{table:varyKE1} and
\ref{table:varyKE2}, rather good estimates of the exact results
can be achieved regardless of how the kinetic energy is
distributed, particularly if one goes to higher order. However,
the best accuracy is obtained if somewhat less kinetic energy
(40\% to be precise) is included in $H_1$, especially for the
first excited state. This result can be understood from
Tables~\ref{table:gamma1} and \ref{table:gamma2} which show that
the perturbation series for $H'(\lambda)$ converges faster
than $H(\lambda)$ to the exact result. This behavior in turn suggests
that for $\gamma = 1$, the linear potential is weaker than the quadratic
potential. Hence, according to the results in Tables~\ref{table:ho} and
\ref{table:af}, a more accurate Pad\'{e} approximant will be
obtained by underweighting the kinetic energy in $H_1$.
Unfortunately, there are no quantitative rules for how the kinetic
energy should be distributed among the two sub-Hamiltonians.
Tables~\ref{table:varyKE1} and \ref{table:varyKE2} suggest that a
$50:50$ split should produce reasonably good estimates.

\section{Conclusion}
The stationary state problems considered here provide a good
illustration of the power of Pad\'{e} interpolation for problems
for which exact solutions are difficult to obtain and ordinary
perturbation methods are not applicable. For practice, the interested
reader may wish to apply the method to interpolate the strong-field
and weak-field Zeeman effects in hydrogen. Exact results for the
$n = 2$ level can be found in Ref.~\onlinecite{bib:Griffiths}.  They
involve square-root functions of the expansion parameter (the magnitude
of the magnetic field), similar to the first example discussed in
Sec.~\ref{sec3}. See also Ref.~\onlinecite{bib:2dHinB} which
discusses the case of the two-dimensional hydrogen atom.

The use of Pad\'{e} interpolation is of course not limited to
quantum mechanical problems, because all that is needed is an
expansion parameter, be it a physical one as in
Eq.~(\ref{eq:Ha>>b}) or an artificial one such as the
interpolation parameter in Eq.~(\ref{eq:interpolatingH}), for
which the behavior of the physical system can be calculated or
measured when the parameter is small as well as when it is large.
We encourage the reader to find other applications of this useful
approximation scheme.

\acknowledgments{This work was supported in part by the U. S.
Department of Energy under grant DE-FG02-84ER40163. We thank
A. Halprin for a discussion, and E. J. Weniger for communicating
to us his work on two-point Pad\'{e} approximants and for calling
our attention to Refs.~\onlinecite{bib:2ptPade} and
\onlinecite{bib:2dHinB}.}

\newpage

\begin{table}[htbp]
 \caption{\label{table:2state}The eigenvalue $E_+$ (in units of
$|a|$) of the two-state system (\ref{eq:2stateH}) for various
values of the parameter $|b/a|$.}
\begin{ruledtabular}
\begin{tabular}{ccc}
$|b/a|$ & Pad\'{e} & Exact
\\ \hline
$0.1$ & $1.00517$ & $1.00499$ \\
$0.5$ & $1.12500$ & $1.11803$ \\
$1$ & $1.42857$ & $1.41421$ \\
$2$ & $2.25000$ & $2.23607$ \\
$10$ & $10.0517$ & $10.0499$ \\
\end{tabular}
\end{ruledtabular}
\end{table}

\newpage

\begin{table}[htbp]
\caption{\label{table:gamma1}Ground state energy, $\gamma =1$.
Columns two to five display the first, second, and third order
perturbative solutions to the Hamiltonians $H(\alpha)$,
$H(\beta)$, $H(\lambda)$, and $H'(\lambda)$, where the
perturbation parameters $\alpha$, $\beta$, $\lambda$, and
$1/\lambda$ are all set to unity. Results from the Pad\'{e}
interpolation of $H(\lambda)$ appear in column six. These are to
be compared with the exact energy, shown in the bottom, obtained
from numerical integration of the Schr\"{o}dinger equation.}
\begin{ruledtabular}
\begin{tabular}{lccccc}
& $H(\alpha)$ & $H(\beta)$ & $H(\lambda)$ & $H'(\lambda)$ & Pad\'{e}
\\ \hline
1st order & $2.62838$ & $2.77411$ & $2.54242$ & $2.53984$ & $2.53724$ \\
2nd order & $2.51908$ & $2.30374$ & $2.52928$ & $2.53517$ & $2.53720$ \\
3rd order & $2.54121$ & $2.88137$ & $2.54998$ & $2.53882$ & $2.53720$ \\
\\
Exact & 2.53720 &&&& \\
\end{tabular}
\end{ruledtabular}
\end{table}

\newpage

\begin{table}[htbp]
\caption{\label{table:gamma2}First excited state energy,
$\gamma=1$. See Table~\protect\ref{table:gamma1} caption for a
detailed description.}
\begin{ruledtabular}
\begin{tabular}{lccccc}
& $H(\alpha)$ & $H(\beta)$ & $H(\lambda)$ & $H'(\lambda)$ & Pad\'{e}
\\ \hline
1st order & $5.19257$ & $6.05194$ & $5.20217$ & $5.13559$ & $5.10483$ \\
2nd order & $5.09417$ & $3.67756$ & $4.90789$ & $5.08365$ & $5.10380$ \\
3rd order & $5.08881$ & $8.19137$ & $5.53588$ & $5.11655$ & $5.10333$ \\
\\
Exact & 5.10382 &&&& \\
\end{tabular}
\end{ruledtabular}
\end{table}

\newpage

\begin{table}[htbp]
\caption{Perturbation theory versus Pad\'{e} interpolation:
 dominant quadratic potential, $\gamma=0.1$. Columns two and three
 show respectively the results from standard perturbative
 calculations with the linear potential as the perturbation, and
 the corresponding estimates from the Pad\'{e} interpolation of
 $H(\lambda)$, where the original kinetic energy term is distributed
 among the sub-Hamiltonians (\ref{eq:h1}) and (\ref{eq:h2}) in
 the ratio $1:9$. The exact energies, obtained from numerically
 integrating the Schr\"{o}dinger equation, are also displayed.
\label{table:ho}}
\begin{ruledtabular}
\begin{tabular}{lcc} & Perturbation Theory & Pad\'{e}
\\ \hline Ground state
\\ 1st order & $1.61284$ & $1.61178$
\\ 2nd order & $1.61174$ & $1.61177$
\\ 3rd order & $1.61177$ & $1.61177$ \\
\\ Exact & $1.61177$
\\ \hline First excited state
\\ 1st order & $3.66926$ & $3.66844$
\\ 2nd order & $3.66827$ & $3.66828$
\\ 3rd order & $3.66827$ & $3.66828$ \\
\\ Exact & $3.66828$ \\
\end{tabular}
\end{ruledtabular}
\end{table}

\newpage

\begin{table}[htbp]
\caption{Perturbation theory versus Pad\'{e} interpolation:
 dominant linear potential, $\gamma=10$. Columns two and three
 show respectively the results from standard perturbative
 calculations with the quadratic potential as the perturbation, and
 the corresponding estimates from Pad\'{e} interpolation of
 $H(\lambda)$, where the original kinetic energy term is distributed
 among the sub-Hamiltonians (\ref{eq:h1}) and (\ref{eq:h2}) in
 the ratio $9:1$. The exact energies, obtained from numerically
 integrating the Schr\"{o}dinger equation, are also displayed.
\label{table:af}}
\begin{ruledtabular}
\begin{tabular}{lcc} & Perturbation Theory & Pad\'{e}
\\ \hline Ground state
\\ 1st order & 8.81152 & 8.80704
\\ 2nd order & 8.80681 & 8.80706
\\ 3rd order & 8.80708 & 8.80706 \\
\\ Exact & 8.80706
\\ \hline First excited state
\\ 1st order & 15.6650 & 15.6432
\\ 2nd order & 15.6412 & 15.6431
\\ 3rd order & 15.6433 & 15.6431 \\
\\ Exact & 15.6431 \\
\end{tabular}
\end{ruledtabular}
\end{table}

\newpage

\begin{table}[htbp]
\caption{Comparisons similar to the last three columns in
Table~\ref{table:gamma1},
 except the original kinetic energy is distributed among the
 sub-Hamiltonians (\ref{eq:h1}) and (\ref{eq:h2}) in the ratios
 indicated.\label{table:varyKE1}}
\begin{ruledtabular}
\begin{tabular}{lccc}
& $H(\lambda)$ & $H'(\lambda)$ & Pad\'{e}
\\ \hline $1:9$
\\ 1st order & 3.64332 & 2.60113 & 2.54126
\\ 2nd order & $-2.96766$ & 2.52016 & 2.53730
\\ 3rd order & 33.5854 & 2.54173 & 2.53721
\\ \hline $2:8$
\\ 1st order & 2.84633 & 2.57650 & 2.53717
\\ 2nd order & 1.66075 & 2.52341 & 2.53736
\\ 3rd order & 5.07816 & 2.54187 & 2.53720
\\ \hline $3:7$
\\ 1st order & 2.62009 & 2.55603 & 2.53703
\\ 2nd order & 2.35777 & 2.52850 & 2.53720
\\ 3rd order & 2.87926 & 2.54110 & 2.53720
\\ \hline $4:6$
\\ 1st order & 2.54956 & 2.54229 & 2.53718
\\ 2nd order & 2.51819 & 2.53396 & 2.53720
\\ 3rd order & 2.56636 & 2.53924 & 2.53720
\\ \hline $5:5$
\\ 1st order & 2.54242 & 2.53984 & 2.53724
\\ 2nd order & 2.52928 & 2.53517 & 2.53720
\\ 3rd order & 2.54998 & 2.53882 & 2.53720
\\ \hline $6:4$
\\ 1st order & 2.56633 & 2.55756 & 2.52772
\\ 2nd order & 2.49717 & 2.51796 & 2.53718
\\ 3rd order & 2.59773 & 2.55661 & 2.53720
\\ \hline $7:3$
\\ 1st order & 2.60713 & 2.61519 & 2.54018
\\ 2nd order & 2.45110 & 2.43389 & 2.53704
\\ 3rd order & 2.66226 & 2.68987 & 2.53721
\\ \hline $8:2$
\\ 1st order & 2.65771 & 2.76705 & 2.54828
\\ 2nd order & 2.40135 & 2.07070 & 2.53623
\\ 3rd order & 2.73243 & 3.65519 & 2.53729
\\ \hline $9:1$
\\ 1st order & $2.71414$ & $3.24341$ & $2.57328$
\\ 2nd order & $2.35175$ & $-0.240897$ & $2.53211$
\\ 3rd order & $2.80562$ & $16.5922$ & $2.53797$\\
\end{tabular}
\end{ruledtabular}
\end{table}
\newpage

\begin{table}[htbp]
\caption{Comparisons similar to the last three columns in
Table~\ref{table:gamma2},
 except the original kinetic energy is distributed among the
 sub-Hamiltonians (\ref{eq:h1}) and (\ref{eq:h2}) in the ratios
 indicated.
\label{table:varyKE2}}
\begin{ruledtabular}
\begin{tabular}{lccc}
& $H(\lambda)$ & $H'(\lambda)$ & Pad\'{e}
\\ \hline $1:9$
\\ 1st order & $6.62887$ & $5.15343$ & $5.10321$
\\ 2nd order & $-3.53137$ & $5.09570$ & $5.10468$
\\ 3rd order & $80.5535$ & $5.09344$ & $5.10330$
\\ \hline $2:8$
\\ 1st order & $5.38749$ & $5.12254$ & $5.10284$
\\ 2nd order & $4.10020$ & $5.09973$ & $5.10380$
\\ 3rd order & $10.1569$ & $5.09757$ & $5.10302$
\\ \hline $3:7$
\\ 1st order & $5.11949$ & $5.10398$ & $5.10341$
\\ 2nd order & $5.05046$ & $5.10423$ & $5.10382$
\\ 3rd order & $5.40290$ & $5.10063$ & $5.10344$
\\ \hline $4:6$
\\ 1st order & $5.11002$ & $5.10443$ & $5.10377$
\\ 2nd order & $5.08677$ & $5.10370$ & $5.10382$
\\ 3rd order & $5.14789$ & $5.10373$ & $5.10382$
\\ \hline $5:5$
\\ 1st order & $5.20217$ & $5.13559$ & $5.10483$
\\ 2nd order & $4.90789$ & $5.08365$ & $5.10380$
\\ 3rd order & $5.53588$ & $5.11655$ & $5.10333$
\\ \hline $6:4$
\\ 1st order & $5.34182$ & $5.21984$ & $5.10989$
\\ 2nd order & $4.66966$ & $5.00364$ & $5.10355$
\\ 3rd order & $6.03944$ & $5.18832$ & $5.04811$
\\ \hline $7:3$
\\ 1st order & $5.50568$ & $5.40620$ & $5.12506$
\\ 2nd order & $4.41718$ & $4.73387$ & $5.10225$
\\ 3rd order & $6.56543$ & $5.57852$ & $5.10083$
\\ \hline $8:2$
\\ 1st order & $5.68231$ & $5.82763$ & $5.16388$
\\ 2nd order & $4.16511$ & $3.73063$ & $5.09690$
\\ 3rd order & $7.09878$ & $8.09921$ & $5.10182$
\\ \hline $9:1$
\\ 1st order & $5.86556$ & $7.03919$ & $5.26810$
\\ 2nd order & $3.91817$ & $-2.12398$ & $5.07466$
\\ 3rd order & $7.63987$ & $39.8902$ & $5.10548$ \\
\end{tabular}
\end{ruledtabular}
\end{table}

\end{document}